%% file: main.tex
\documentclass[conference]{IEEEtran}
\usepackage{cite}
\usepackage{amsmath,amssymb,amsfonts}
\usepackage{algorithmic}
\usepackage{graphicx}
\usepackage{textcomp}
\usepackage{xcolor}
\usepackage{url}
\usepackage{capt-of}    
\usepackage[hidelinks]{hyperref} 
\usepackage{booktabs}

\def\BibTeX{{\rm B\kern-.05em{\sc i\kern-.025em b}\kern-.08em
    T\kern-.1667em\lower.7ex\hbox{E}\kern-.125emX}}
\begin{document}

\title{Feel My Focus: Gaze-Triggered Digital Twin Creation for Remote Collaboration}

\author{
\IEEEauthorblockN{Shih-Yu Ma}
\IEEEauthorblockA{\textit{Department of Computer Science}\\
\textit{University of Colorado Boulder}\\
Boulder, CO, USA\\
shih-yu.ma@colorado.edu}
\and
\IEEEauthorblockN{Suibi Che-Chuan Weng}
\IEEEauthorblockA{\textit{ATLAS Institute}\\
\textit{University of Colorado Boulder}\\
Boulder, CO, USA\\
chwe1250@colorado.edu}
\and
\IEEEauthorblockN{Ming-Yuan Pai}
\IEEEauthorblockA{\textit{Department of Music}\\
\textit{National Taiwan Normal University}\\
Taipei, Taiwan\\
maggie.myp.music@gmail.com}
\and
\IEEEauthorblockN{Tom Yeh}
\IEEEauthorblockA{\textit{Department of Computer Science}\\
\textit{University of Colorado Boulder}\\
Boulder, CO, USA\\
tom.yeh@colorado.edu}
\and
\IEEEauthorblockN{Ellen Yi-Luen Do}
\IEEEauthorblockA{\textit{ATLAS Institute}\\
\textit{University of Colorado Boulder}\\
Boulder, CO, USA\\
ellen.do@colorado.edu}
\and
\IEEEauthorblockN{}
\IEEEauthorblockA{}
}

\maketitle

\begin{abstract}
\input{sections/00_abstract}

\end{abstract}

\begin{IEEEkeywords}
telepresence, mixed reality, eye tracking, one-shot 3D reconstruction, remote collaboration, digital twins
\end{IEEEkeywords}


\par\smallskip
\noindent
\begin{minipage}{\columnwidth}
\footnotesize

\noindent
\textcopyright{} 2026 IEEE. Personal use of this material
is permitted. Permission from IEEE must be obtained for
all other uses, in any current or future media, including
reprinting/republishing this material for advertising or
promotional purposes, creating new collective works,
for resale or redistribution to servers or lists, or
reuse of any copyrighted component of this work in
other works.

\medskip

\noindent
Published in the 2026 IEEE International Conference
on Artificial Intelligence and eXtended and Virtual
Reality (AIxVR), pp. 295--299, 2026.

\noindent
DOI: \href{https://doi.org/10.1109/AIxVR67263.2026.00052}%
{10.1109/AIxVR67263.2026.00052}

\end{minipage}
\par\smallskip

\input{sections/01_introduction}

\input{sections/02_related_work}

\input{sections/03_system_design}

\input{sections/04_implementation}

\input{sections/05_demo}

\input{sections/06_discussion_future_work}

\bibliographystyle{IEEEtran}
\bibliography{reference}

\end{document}

%% file: sections/00_abstract.tex
We present \emph{Feel My Focus}, a gaze-triggered mixed reality (MR) demonstration that enables rapid creation and sharing of lightweight 3D digital twins during remote collaboration. Unlike systems that use gaze primarily as a referential pointer, Feel My Focus treats sustained visual attention as a generative action: a fixation on a physical object initiates single-image capture, attention-guided segmentation, and one-shot 3D reconstruction, allowing a remote partner to inspect and discuss the object in situ.

The system integrates gaze-tracking glasses, modular segmentation and reconstruction services, and an MR client for runtime streaming and interaction. This design prioritizes responsiveness and experiential continuity over photometric fidelity. We report observations from a public demonstration, highlighting design trade-offs, failure modes, and user control considerations for deploying gaze-triggered generative systems in real-world collaborative settings.


%% file: sections/01_introduction.tex
\section{Introduction}

Imagine a long-distance couple on a video call. One partner at a museum wants to share a small statue—not as a flat photo, but as something the other can immediately pick up, rotate, and discuss in mixed reality (MR). Contemporary tools connect faces and voices well, yet they fail to convey the \emph{things} that carry personal meaning. Moments like these raise a simple question: how can a person instantly recreate and share a meaningful object so that someone else, elsewhere, can experience it as if present?

\begin{figure}[t]
  \centering
  \includegraphics[width=\linewidth]{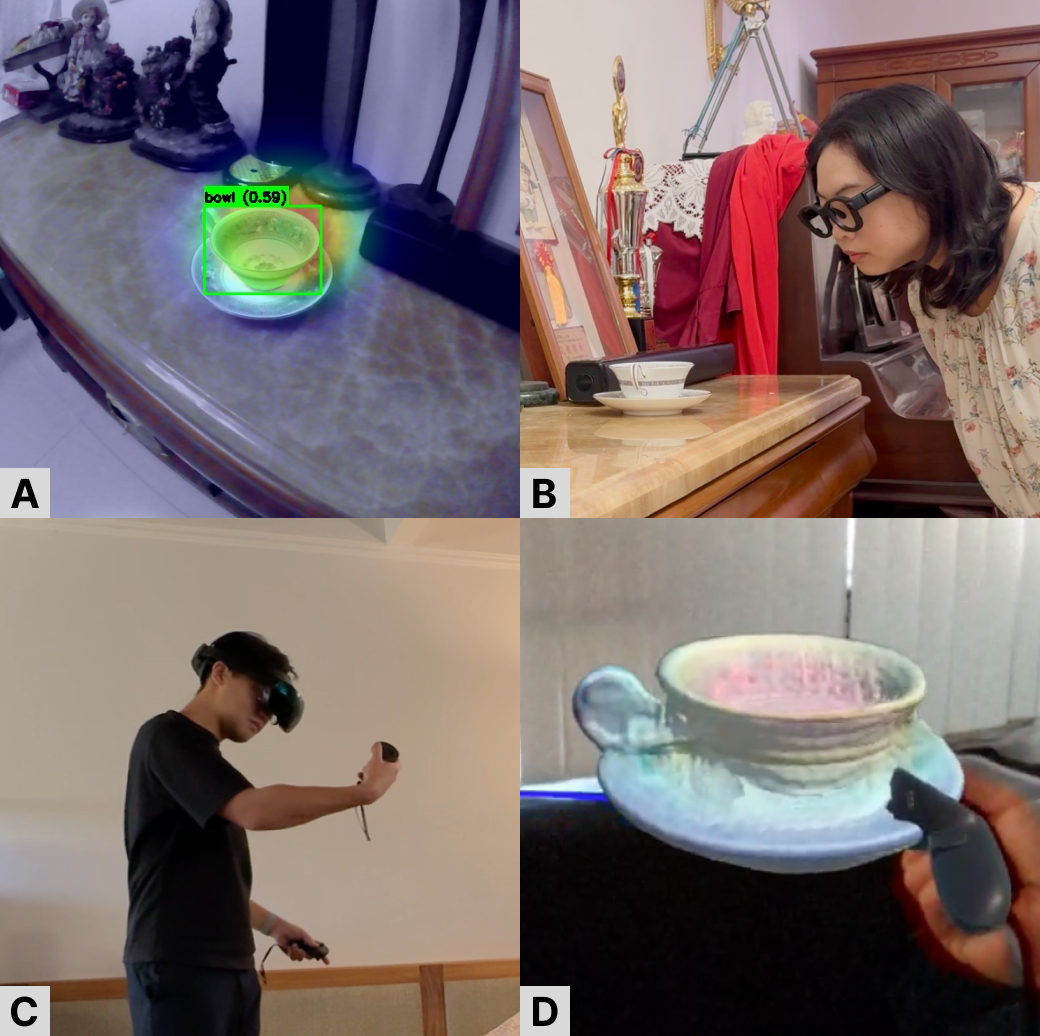}
  \caption{%
  Overview of \textbf{Feel My Focus}. 
  (A) The Capture User’s gaze highlights an attended object, which is detected and segmented. 
  (B) The Capture User wearing gaze-tracking glasses focuses on a real-world item. 
  (C) The MR User receives the reconstructed object in an immersive environment. 
  (D) The digital twin appears in the MR scene for inspection and interaction.
  }
  \label{fig:hero}
\end{figure}

We present \textbf{Feel My Focus}, a system that treats gaze as a \emph{generative trigger} for digital-twin creation. Sustained attention on a real object initiates a single-view reconstruction that is streamed to a remote MR client, turning “what I am looking at now” into “what we can handle together.” Unlike prior work that uses gaze primarily as a referential pointer, Feel My Focus uses attention itself to instantiate shareable content. To support in-conversation use, the system prioritizes \emph{responsiveness over photometric perfection}: rather than relying on multi-view photogrammetry and minutes of processing, it combines intent-triggered capture, segmentation, and one-shot 3D reconstruction to deliver an interactable proxy within a single conversational exchange (Fig.~\ref{fig:hero}). Practical defaults—brief dwell thresholds, immediate progress feedback, lightweight meshes, and runtime asset replacement—ensure that the pipeline remains usable in public and ad-hoc settings.

This work takes a systems and interaction perspective on gaze-triggered generative sharing. We describe the architecture and implementation choices that enable attention-driven digital-twin creation, and we report qualitative observations from real-world demonstrations that reveal design trade-offs related to responsiveness, robustness, user control, and privacy. Together, these elements position Feel My Focus as a live MR demonstration of how gaze can function not only as an awareness cue, but as a mechanism for instantiating shared artifacts during remote collaboration.

This work contributes:
\begin{itemize}
  \item A live MR demonstration that reframes sustained gaze from a referential cue into a generative trigger for instantiating shareable 3D digital twins.
  \item An end-to-end, low-latency pipeline enabling in-conversation object sharing through gaze-guided capture, segmentation, and single-image 3D reconstruction.
  \item Qualitative observations from a public deployment, identifying design trade-offs, robustness issues, and interaction constraints for gaze-triggered generative systems.
\end{itemize}

%% file: sections/02_related_work.tex
\section{Related Work}

Our work draws from three intersecting areas: (A) telepresence and remote collaboration, (B) eye tracking in XR, and (C) generative 3D reconstruction. Together, these strands motivate how gaze can function as both an interaction modality and a channel for emotional connection in remote collaboration.

\subsection{Telepresence and Remote Collaboration}
Volumetric telepresence systems demonstrated that transmitting high-quality human geometry in real time can strongly support social \textit{co-presence}, but they require multi-camera rigs and substantial computational resources (e.g., Microsoft’s \textit{Holoportation} prototype at UIST 2016)~\cite{orts2016holoportation}. Subsequent XR remote-collaboration research shifted toward sharing task context and attention cues rather than full-scene capture. Bai et al.\ showed that combining a live 3D panorama with shared gaze and hand gestures improves co-presence and spatial communication compared to gaze alone~\cite{bai2020mrCollab}. Wang et al.\ compared head-pointer and eye-gaze visualization for spatial AR guidance, demonstrating that gaze serves as an effective referential cue in physical tasks~\cite{wang2020headVsEye}. Surveys and design studies further map the design space of XR remote collaboration, highlighting visual communication cues and shared views as key levers for performance and mutual understanding~\cite{wang2021armrReview,schafer2022survey}. 

In contrast to systems that transmit people or scenes, Feel My Focus centers on sharing \textit{objects of attention}. By turning gaze into a capture trigger that produces a lightweight digital twin, the system preserves the situatedness of attention and supports mediated co-presence through meaningful artifacts such as gifts or personal items.

\subsection{Eye Tracking in XR}
Gaze is a fast, low-effort modality that reveals intention and joint attention. In co-located AR, Jing et al.\ showed that bi-directional gaze visualizations increase perceived joint attention and engagement compared to non-collaborative baselines~\cite{jing2021bidirectional}. In remote MR, sharing gaze behaviors enhances co-presence and coordination, particularly when combined with other cues~\cite{bai2020mrCollab,bai2022gazeBehaviors}. A recent study comparing hand and gaze pointers reported benefits for message understanding, task load, and user preference, especially when cues are combined~\cite{shin2024handGaze}. Beyond pointing, platforms that integrate gaze with intermediate ``2.5D'' depth cues support practical assembly and training workflows~\cite{wang2022twoPointFiveD}.

Prior work has also used gaze to guide recognition and segmentation. Toyama et al.~\cite{toyama2012gaze} demonstrated early use of head-mounted eye tracking to constrain object recognition to regions of visual attention. More recently, Wang et al.~\cite{wang2024gazesam} introduced GazeSAM, which combines eye gaze with the Segment Anything Model to support interactive image segmentation. These systems use gaze to disambiguate user intent or refine selection within an image.

Feel My Focus adopts a different interaction framing. Rather than guiding perception or selection alone, sustained gaze initiates the creation of a shareable artifact. The system treats attention as a generative trigger that instantiates a lightweight 3D proxy for remote mixed reality collaboration.

\begin{figure*}[ht]
    \centering
    \includegraphics[width=\textwidth]{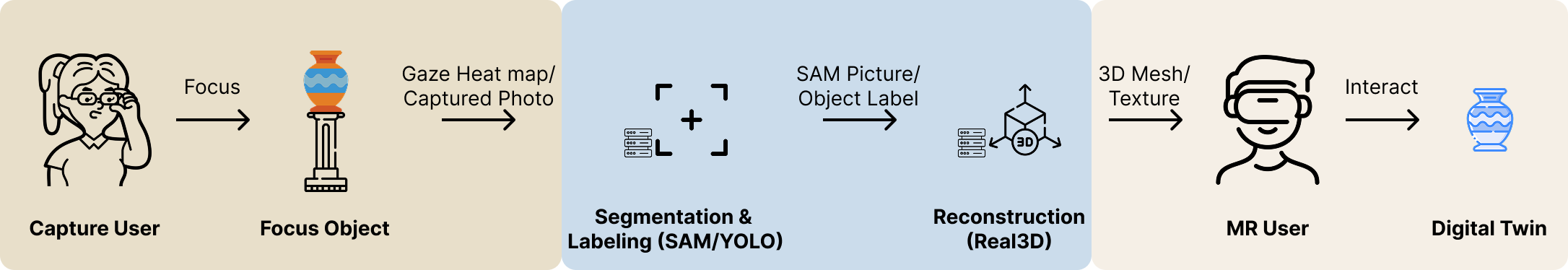}
    \caption{Pipeline of \textbf{Feel My Focus}. A sustained gaze fixation from the Capture User is captured through gaze-tracking glasses and processed into a heatmap and image. The SAM/YOLO service segments the attended object and assigns a label, which is then reconstructed into a textured 3D mesh by Real3D. The resulting digital twin is transmitted to the MR User for immersive inspection and interaction.}
    \label{fig:pipeline}
\end{figure*}

\subsection{Generative 3D Reconstruction: Responsiveness vs.\ Fidelity}
Classical photogrammetry pipelines such as \textit{AliceVision Meshroom}~\cite{alicevision2021} achieve high-quality 3D models through structure-from-motion and multi-view stereo reconstruction. These approaches produce photorealistic meshes with accurate geometry and texture, but they require dozens of images captured from multiple viewpoints, careful calibration, and substantial processing time. In contrast, Feel My Focus prioritizes responsiveness: a single gaze-triggered capture produces a shareable digital twin suitable for in-conversation use.

Recent diffusion-based pipelines further advance single-image 3D reconstruction. \textit{Zero-1-to-3} conditions a diffusion prior on camera transforms to synthesize consistent novel views~\cite{liu2023zero123}. \textit{RealFusion} combines a single input with a diffusion prior while fitting a neural radiance field (NeRF), producing plausible 360° reconstructions of unseen surfaces~\cite{melas2023realfusion}. Feed-forward systems such as \textit{One-2-3-45} and its successor reduce latency to seconds while improving multi-view consistency~\cite{liu2023one2345,liu2024one2345pp}. Although text-to-3D approaches (e.g., \textit{Magic3D}) target prompt-based content creation~\cite{lin2023magic3d}, they rely on similar generative priors.

By favoring responsiveness over photometric perfection, Feel My Focus transforms a moment of gaze into a rapidly generated digital twin. This design supports attention-driven communication in time-sensitive collaborative settings, while positioning higher-fidelity reconstruction as a complementary direction for future work.

%% file: sections/03_system_design.tex
\section{System Design}

\textbf{Feel My Focus} enables gaze-driven creation of digital twins through a modular, multi-service pipeline (Fig.~\ref{fig:pipeline}). The system defines two user roles: the \emph{Capture User}, who wears gaze-tracking glasses, and the \emph{MR User}, who receives and interacts with reconstructed objects in an immersive environment. These roles are not fixed; users can switch between capturing and viewing depending on the collaborative context.

\subsection{Architecture Overview}
The architecture separates sensing, recognition, reconstruction, and visualization into distinct services coordinated by an orchestrator. This modular structure allows components to operate independently while supporting real-time communication across devices. In this system, gaze functions as a generative trigger: sustained attention initiates capture, which is processed into a digital twin and shared in mixed reality. By coupling gaze-driven capture with one-shot 3D reconstruction, the pipeline prioritizes responsiveness and co-presence over full-scene fidelity.

\subsection{Workflow}
The workflow proceeds through the following stages:
\begin{enumerate}
  \item \textbf{Focus Capture} — The system tracks the Capture User’s gaze and records an image aligned with the attended region.
  \item \textbf{Attention-Guided Recognition} — The captured image is segmented and labeled to isolate the attended object.
  \item \textbf{Selection} — Candidate objects are presented to the MR User, who confirms the target for reconstruction.
  \item \textbf{Digital Twin Generation} — A one-shot 3D reconstruction service generates a textured mesh of the selected object.
  \item \textbf{Immersive Interaction} — The mesh is transmitted to the MR User’s client, where it can be inspected, manipulated, and placed within the mixed reality scene.
\end{enumerate}

Together, the architecture and workflow describe how gaze input is transformed into a shareable digital twin, providing the foundation for the implementation details presented next.

%% file: sections/04_implementation.tex
\section{Implementation}

The prototype comprises three components: a wearable capture device, a server infrastructure, and a mixed reality (MR) client.

\subsection{Wearable Capture Device}
The system uses \textbf{GANZIN Sol Glasses}\footnote{\url{https://ganzin.com/en/sol-glasses-wearable-eye-tracker/}}, which integrate binocular eye tracking with a forward-facing RGB camera. The client follows a two-stage dwell mechanism aligned with Microsoft’s gaze-interaction guidelines~\cite{MSFT_GazeDwell}. An initial 250\,ms onset filters incidental glances and displays a subtle circular cue. If fixation persists for an additional 750\,ms (approximately 1.0\,s total), the system commits to capture and enters a gaze-tracking phase. During this phase, a thin progress indicator is shown while gaze samples are recorded for 3.0\,s to compose a fixation heatmap. At the end of this window, the client captures an RGB frame and transmits the image and heatmap to the server. If gaze leaves the target before dwell completion, capture is canceled and no data are transmitted. Raw frames are processed locally and discarded after heatmap generation.

\subsection{Server Infrastructure}
Segmentation and reconstruction services run on a high-performance laptop (Intel Core i7-14700HX CPU, NVIDIA RTX 4070 GPU, 64\,GB RAM). The infrastructure consists of two services connected via TCP sockets to both the capture device and the MR client:
\begin{itemize}
    \item \textbf{Segmentation and Recognition} — A SAM/YOLO pipeline~\cite{kirillov2023segmentAnything,yolov8_ultralytics} combines the gaze heatmap with object detection to isolate the attended region and assign candidate labels.
    \item \textbf{Reconstruction} — A Real3D service~\cite{jiang2024real3d} receives the segmented image and performs single-image 3D reconstruction, producing a textured mesh suitable for real-time rendering.
\end{itemize}
This modular organization allows services to operate independently, supports substitution of alternative back ends, and maintains reliable low-latency communication across devices.

\subsection{MR Visualization Client}
The MR client runs on a \textbf{Meta Quest 3} headset and is implemented in Unity to support runtime asset streaming and interaction. The headset connects to the server via TCP to receive object data. When the Capture User fixates on an object, the MR client receives candidate labels and presents them to the MR User for confirmation. After selection, the reconstructed mesh is streamed into the MR scene, where it can be inspected, manipulated, and placed. Prior work has established the feasibility of runtime asset integration in Unity~\cite{giunchi2024dreamcodevr,jennings2024gromit,delatorre2023llmr}; this implementation extends these capabilities to gaze-triggered digital twins with progressive streaming from a coarse proxy to a refined mesh.

\subsection{Representative Demo Outputs}
\begin{figure}[t]
  \centering
  \includegraphics[width=0.8\linewidth]{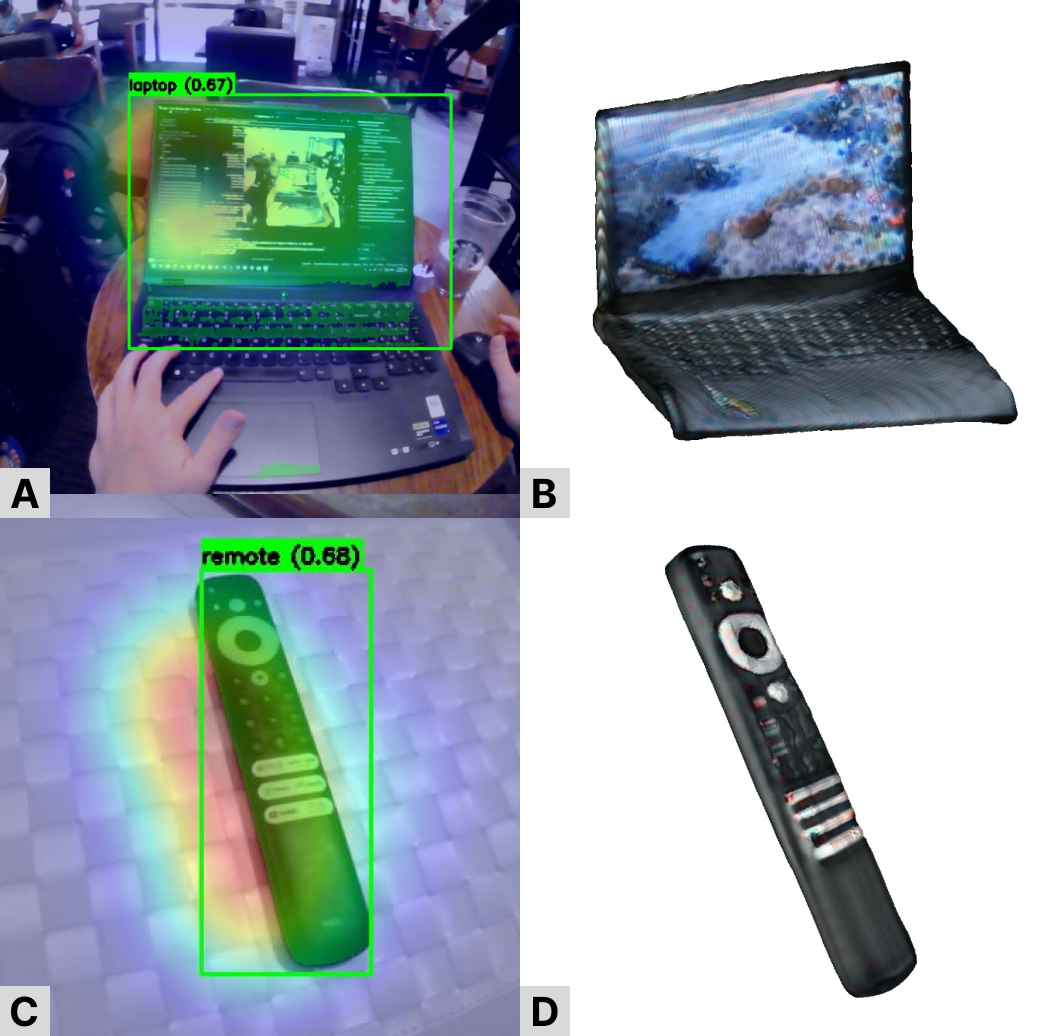}
  \caption{Representative demo outputs of Feel My Focus. (A) Gaze-guided capture with attention heatmap and detection box for a laptop, and (B) the resulting digital twin streamed to the MR client. (C--D) Corresponding capture and reconstructed digital twin for a handheld remote control. These examples illustrate how sustained gaze instantiates shareable 3D proxies for live mixed reality interaction.}
  \label{fig:laptop-remote}
\end{figure}

Figure~\ref{fig:laptop-remote} presents representative outputs from live demonstrations of Feel My Focus. In these examples, fixation on a laptop or a handheld remote control initiates gaze-guided capture, segmentation, and single-image 3D reconstruction, resulting in a lightweight digital twin streamed to the MR client. Reconstruction quality varies with object geometry, texture, and reflectance, yet the generated meshes consistently support inspection, manipulation, and conversational reference in mixed reality. These examples demonstrate the system’s ability to instantiate shareable 3D proxies in situ, prioritizing immediacy and interaction continuity over photometric fidelity.

%% file: sections/05_demo.tex
\section{Demo Observations and Failure Modes}

We demonstrated Feel My Focus in a public exhibition setting, where 35 visitors interacted with the system for approximately 5--8 minutes each. The observations reported here reflect behavior and system performance during live deployment.

Participants consistently prioritized immediacy over reconstruction fidelity, valuing the ability to share and manipulate objects within a single conversational turn. Common failure modes included glossy or low-texture objects, cluttered backgrounds that led to ambiguous segmentation, and occasional false gaze triggers. When detection confidence was low, reconstruction quality degraded gracefully yet remained usable as a conversational proxy.

To maintain robustness during live interaction, generated assets can be rapidly replaced or discarded without interrupting the interaction flow. These observations directly informed system defaults and illustrate practical trade-offs involved in deploying gaze-triggered generative systems in real-world collaborative settings.

%% file: sections/06_discussion_future_work.tex
\section{Discussion and Future Work}

The prototype demonstrates that gaze can instantiate shareable digital twins during live mixed reality interaction, while also revealing practical constraints inherent to real-world deployment. These constraints inform directions for refinement rather than shortcomings of the demonstration itself.

\textbf{Performance and scalability.} The current implementation prioritizes interaction continuity over detailed system profiling. Future iterations will characterize end-to-end latency and bandwidth usage across common network conditions (e.g., Wi-Fi and 5G) to guide deployment at larger scales.

\textbf{Selection robustness.} Trigger reliability and segmentation quality degrade in scenes with overlapping objects, reflective or transparent surfaces, and low-texture materials. Shorter dwell thresholds combined with optional lightweight confirmation mechanisms, as well as improved class-agnostic masking, represent concrete paths to increase robustness without increasing interaction burden.

\textbf{User control and safeguards.} While gaze-only triggering simplifies interaction during live demonstrations, optional confirmation steps can increase user agency and reduce unintended capture. Integrating such controls alongside clear visual indicators and local processing safeguards will be important for broader deployment.

Together, these directions support the evolution of Feel My Focus from an exploratory demonstration toward more robust, user-controlled, and scalable gaze-triggered collaborative systems.

\section{Conclusion}

We presented \textbf{Feel My Focus}, a system that reimagines gaze as a \emph{generative input} rather than a passive indicator. Unlike prior telepresence approaches that rely on multi-view capture or pre-modeled assets, Feel My Focus demonstrates how sustained gaze can instantiate shareable content during live mixed reality collaboration, rather than merely pointing to it. By coupling gaze-triggered capture with single-image 3D reconstruction, the system prioritizes responsiveness and co-presence over photometric perfection.

This work positions attention itself as a communicative act: what a person looks at becomes an object that others can inspect, manipulate, and discuss together. As a live demonstration, Feel My Focus illustrates how lightweight, attention-driven digital twins can support new forms of mediated co-presence and object-centered communication across distance.